\documentclass{article}

\usepackage{PRIMEarxiv}

\usepackage[utf8]{inputenc} 
\usepackage[T1]{fontenc}    
\usepackage{hyperref}       
\usepackage{url}            
\usepackage{booktabs}       
\usepackage{amsfonts}       
\usepackage{nicefrac}       
\usepackage{microtype}      
\usepackage{lipsum}
\usepackage{fancyhdr}       
\usepackage{graphicx}       

\usepackage{multirow}
\usepackage{cite}
\usepackage{amsmath,amssymb}
\usepackage{algorithmic}
\usepackage{algorithm}
\usepackage{textcomp}

\hypersetup{hidelinks}

\title{Explainable Interictal Epileptiform Discharge Detection Method Based on Scalp EEG and Retrieval-Augmented Generation

}

\author{
  Yu Zhu$^{1}$ \\
  Nanjing University of Chinese Medicine \\
  Nanjing 210023, China \\
  \And
  Jiayang Guo$^{1}$ \\
  National Institute for Data Science in Health and Medicine \\
  Xiamen University, Xiamen 361005, China \\
  \And
  Jun Jiang, Xin Shu$^{*}$ \\
  Wuhan Children's Hospital \\
  Wuhan 430010, China \\
  \texttt{shuxin@zgwhfe.com} \\
  \And
  Peipei Gu$^{*}$ \\
  Zhengzhou University of Light Industry \\
  Zhengzhou 450002, China \\
  \texttt{gupeipei@zzuli.edu.cn} \\
  \And
  Duo Chen$^{*}$ \\
  Nanjing University of Chinese Medicine \\
  Nanjing 210023, China \\
  \texttt{380013@njucm.edu.cn} \\
}

\begin{document}
\maketitle

\begingroup
\renewcommand\thefootnote{}
\footnotetext{
$^{1}$These authors contributed equally to this work. \protect\\
This work was supported by the National Natural Science Funds of China (62006100) and the Xiamen Natural Science Foundation (3502Z202473010). (*Corresponding authors: Duo Chen; Peipei Gu; Xin Shu) \protect\\
Yu Zhu and Duo Chen are with the School of Artificial Intelligence and Information Technology, Nanjing University of Chinese Medicine, Nanjing 210023, China (e-mail: 380013@njucm.edu.cn). \protect\\
Jiayang Guo is with the National Institute for Data Science in Health and Medicine, Xiamen University, Xiamen 361005, China. \protect\\
Jun Jiang and Xin Shu are with the Department of Electrophysiology, Wuhan Children's Hospital, Tongji Medical College, Huazhong University of Science and Technology, Wuhan 430010, China (e-mail: shuxin@zgwhfe.com). \protect\\
Peipei Gu is with the College of Software Engineering, Zhengzhou University of Light Industry, Zhengzhou 450002, China (e-mail: gupeipei@zzuli.edu.cn).
}
\endgroup

\begin{abstract}
The detection of interictal epileptiform discharge (IED) is crucial for the diagnosis of epilepsy, but automated methods often lack interpretability. This study proposes IED-RAG, an explainable multimodal framework for joint IED detection and report generation. Our approach employs a dual-encoder to extract electrophysiological and semantic features, aligned via contrastive learning in a shared EEG--text embedding space. During inference, clinically relevant EEG--text pairs are retrieved from a vector database as explicit evidence to condition a large language model (LLM) for the generation of evidence-based reports.  Evaluated on a private dataset from Wuhan Children's Hospital and the public TUH EEG Events Corpus (TUEV), the framework achieved balanced accuracies of 89.17\% and 71.38\%, with BLEU scores of 89.61\% and 64.14\%, respectively. The results demonstrate that retrieval of explicit evidence enhances both diagnostic performance and clinical interpretability compared to standard black-box methods.
\end{abstract}


\section{Introduction}
\label{sec:introduction}
Epilepsy, a prevalent neurological disorder affecting over 50 million people worldwide, poses a significant global health burden \cite{ref1}. Its diagnosis and management critically rely on the identification of interictal epileptiform discharge (IED), key electrophysiological biomarkers captured via electroencephalography (EEG) \cite{ref2, ref3}. In clinical practice, electroencephalography (EEG) is the primary modality for capturing these transient patterns across diverse settings, from routine examinations to long-term monitoring. However, manual EEG interpretation is a labor-intensive and cognitively demanding task \cite{ref4}. Neurophysiologists must meticulously review  large volumes of multichannel data to identify sparse and variable epileptiform patterns. This process is often fraught with significant inter- and intra-rater variability \cite{ref4, ref5}. This creates a pressing clinical need for automated tools that not only detect IED accurately but also provide interpretable outputs to support diagnostic decisions \cite{ref5, ref6}.

To address this need, automated IED detection has been a long-standing goal. Early approaches depended on handcrafted feature extraction coupled with traditional machine learning classifiers, such as support vector machines, which often struggled with generalization across patients and EEG non-stationarity \cite{ref7}. The advent of deep learning (DL) has markedly advanced the field. Models based on convolutional neural networks (CNNs), recurrent neural networks (RNNs), and transformers now approach human-level sensitivity in detecting IED by learning features directly from raw signals \cite{ref8}. Despite these performance gains, a critical barrier to clinical adoption remains: most DL models operate as ``black boxes,'' providing detection decisions without clinically interpretable explanations \cite{ref9}. While post hoc explainability techniques like saliency maps can visualize signal regions influencing model decisions \cite{ref10}, these methods typically offer only localization cues, failing to translate complex electrophysiological features into the structured, narrative reports required for clinical communication \cite{ref11}.

Recent breakthroughs in large language models (LLMs), such as GPT and Llama, offer a promising pathway to address this interpretability gap \cite{ref12}. Their exceptional capability in natural language understanding and generation has catalyzed interest in EEG-to-text translation and automated reporting \cite{ref13,ref14}. Emerging research now integrates EEG encoders with a pretrained LLM to generate descriptive narratives directly from physiological signals \cite{ref15, ref16}. This paradigm shifts the output modality from abstract numerical probabilities to clinically actionable statements, such as ``high-amplitude sharp waves observed in the left temporal region.'' This alignment brings AI outputs closer to standard medical reporting workflows \cite{ref17}. However, current EEG-to-text generation models face a critical challenge: ``hallucinations''---the generation of plausible-sounding but factually inaccurate descriptions due to a lack of grounded evidence \cite{ref18}. Without access to explicit clinical precedents, these models often produce reports that are repetitive, overly generic, or insufficiently substantiated by the underlying signal morphology \cite{ref19}. In contrast, human experts rarely interpret EEGs in isolation. Instead, clinicians rely heavily on case-based reasoning (CBR), evaluating current findings against a mental repository of diagnostically confirmed cases \cite{ref20}. This comparative process allows clinicians to contextualize ambiguous patterns, assess similarities to known phenotypes, and support diagnoses with experiential evidence \cite{ref21}. An effective AI system should emulate this cognitive process by explicitly referencing relevant historical cases, ensuring that both detection and explanation are grounded in validated clinical examples rather than purely statistical associations \cite{ref22, ref23}.

Retrieval-Augmented Generation (RAG) has emerged as a powerful framework to mitigate hallucinations by conditioning generative models on retrieved external evidence \cite{ref24}.  It has proven effective in enhancing factual consistency in NLP and vision tasks \cite{ref25}. For instance, retrieved visual prototypes have successfully supported explainable image classification and captioning \cite{ref26}. This paradigm is particularly suited for EEG interpretation, where historical databases of recordings and expert reports constitute a rich, structured knowledge base \cite{ref27}. By retrieving and conditioning on clinically relevant EEG-report pairs, a RAG framework can provide explicit evidentiary support for both detection and narrative generation, closely mirroring the clinician's CBR process \cite{ref28,ref29}.

Motivated by the need for evidence-based and interpretable automation, this study proposes IED-RAG, a multimodal Retrieval-Augmented Generation framework for joint IED detection and clinical report generation. The framework first aligns raw EEG segments with expert textual descriptions in a shared embedding space through contrastive learning, utilizing a Deep4Net-based encoder for spatiotemporal EEG features and a BERT-base-uncased encoder for text. During inference, a FAISS-accelerated vector database retrieves the most clinically similar historical IED cases. This retrieved evidence explicitly conditions a Meta-Llama-3-8B-Instruct large language model to generate diagnostic reports. Evaluated on a private dataset from Wuhan Children's Hospital and the public TUH EEG Events Corpus (TUEV), the framework achieved balanced accuracies of 89.17\% and 71.38\%, respectively. By grounding detection and narrative in retrieved clinical evidence, IED-RAG bridges quantitative signal analysis with qualitative reporting, providing a transparent tool to enhance diagnostic efficiency and foster clinical trust.

   To summarize, our main contributions are twofold:
\begin{itemize}
	\item We propose IED-RAG, a novel multimodal RAG framework that integrates contrastive EEG-text alignment with evidence-retrieval to jointly perform IED detection and explainable report generation.
	\item The framework operationalizes clinical reasoning by retrieving and conditioning on similar, confirmed historical cases, thereby mitigating hallucinations. Its effectiveness and enhanced interpretability are empirically demonstrated on both private and public datasets, showing competitive performance over standard methods.
	
\end{itemize}

  The rest of this paper is organized as follows: Section~\ref{sec:problem} formulates the problem and details the datasets used in this study. Section~\ref{sec:method} describes the proposed IED-RAG methodology. Section~\ref{sec:results} presents the experimental results. Section~\ref{sec:discussion} provides a comprehensive discussion of the findings. Finally, Section~\ref{sec:conclusion} concludes the paper and proposes directions for future work.

\begin{table*}[t]
\centering
\caption{Clinical Profiles, EEG Statistics, and Textual Annotations of Dataset~1 (Wuhan Private Dataset)}
\label{tab:dataset1_profile}
\setlength{\tabcolsep}{8pt}
\renewcommand{\arraystretch}{1.2}
\resizebox{\textwidth}{!}{
\begin{tabular}{c c c c c l c c}
\toprule
\textbf{ID} & \textbf{Sex} & \textbf{Age} & \textbf{Epilepsy Type} & \textbf{IED Location} & \textbf{Key Report Description} & \textbf{\#IED} & \textbf{EEG Duration (s)} \\
\midrule
1  & M & 10 & BECTS & Left Rolandic & Centrotemporal spikes, sleep-activated & 135 & 487 \\
2  & F & 4  & BECTS & Right Rolandic & Frequent sharp waves during sleep & 90  & 1745 \\
3  & M & 11 & Febrile Seizure & Bilateral Occipital & Occipital spikes with slow waves & 124 & 1798 \\
4  & M & 10 & BECTS & Left Rolandic & Repetitive centrotemporal discharges & 55  & 780 \\
5  & F & 5  & BECTS & Bilateral Central & High-amplitude spikes, sleep-related & 131 & 945 \\
6  & M & 4  & Non-epileptic & Bilateral Posterior & No clear epileptiform discharge & 0   & 913 \\
7  & F & 12 & Focal Epilepsy & Bilateral Frontal & Frontal sharp transients & 89  & 3445 \\
8  & M & 6  & Childhood Absence & Generalized & Generalized spike-and-wave complexes & 117 & 3723 \\
9  & M & 10 & Focal Epilepsy & Central Midline & Clustered central spikes & 114 & 1893 \\
10 & M & 9  & Focal Epilepsy & Left Temporal & Temporal spikes during sleep & 111 & 2299 \\
11 & M & 12 & Focal Epilepsy & Right Parietal & Multifocal spikes, posterior dominant & 131 & 932 \\
\bottomrule
\end{tabular}
}
\end{table*}

\begin{figure*}[t]
\centering
\includegraphics[width=\textwidth]{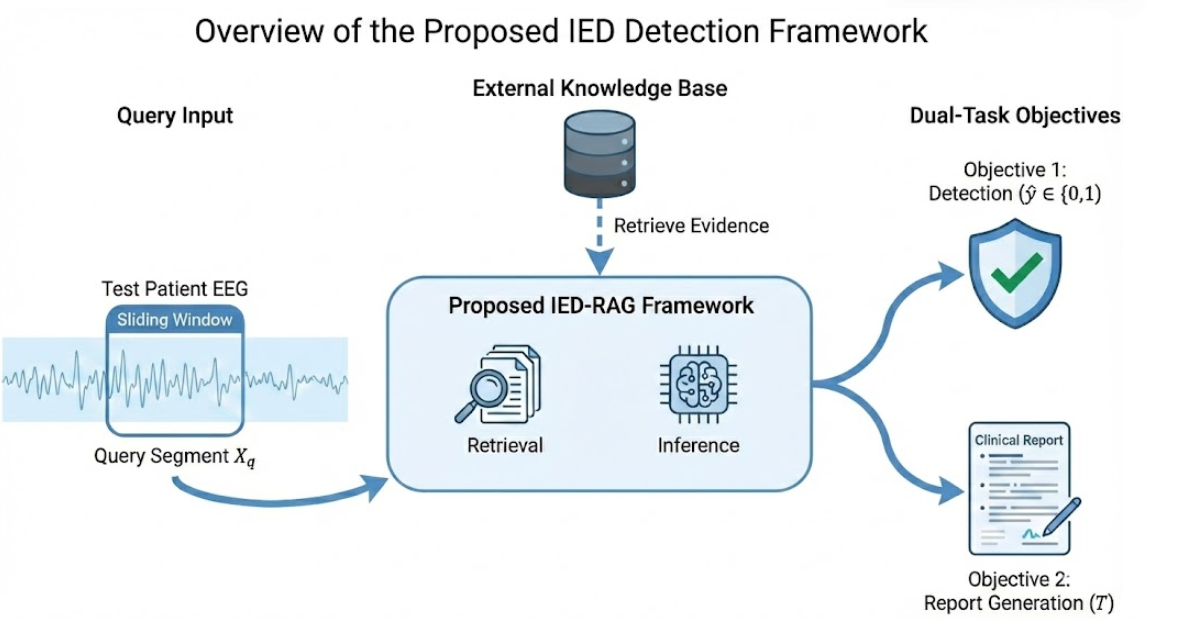}
\caption{Overview of the Proposed IED Detection Framework. The workflow illustrates the evidence-grounded analysis process: A query EEG segment $X_q$ is input into the system. The core IED-RAG framework retrieves clinically relevant evidence from an external knowledge base to ground the inference. The system produces dual outputs: a binary detection decision $\hat{y}$ and an interpretable clinical report $T$.}
\label{fig:task_overview}
\end{figure*}

\begin{figure*}[t]
\centering
\includegraphics[width=\textwidth]{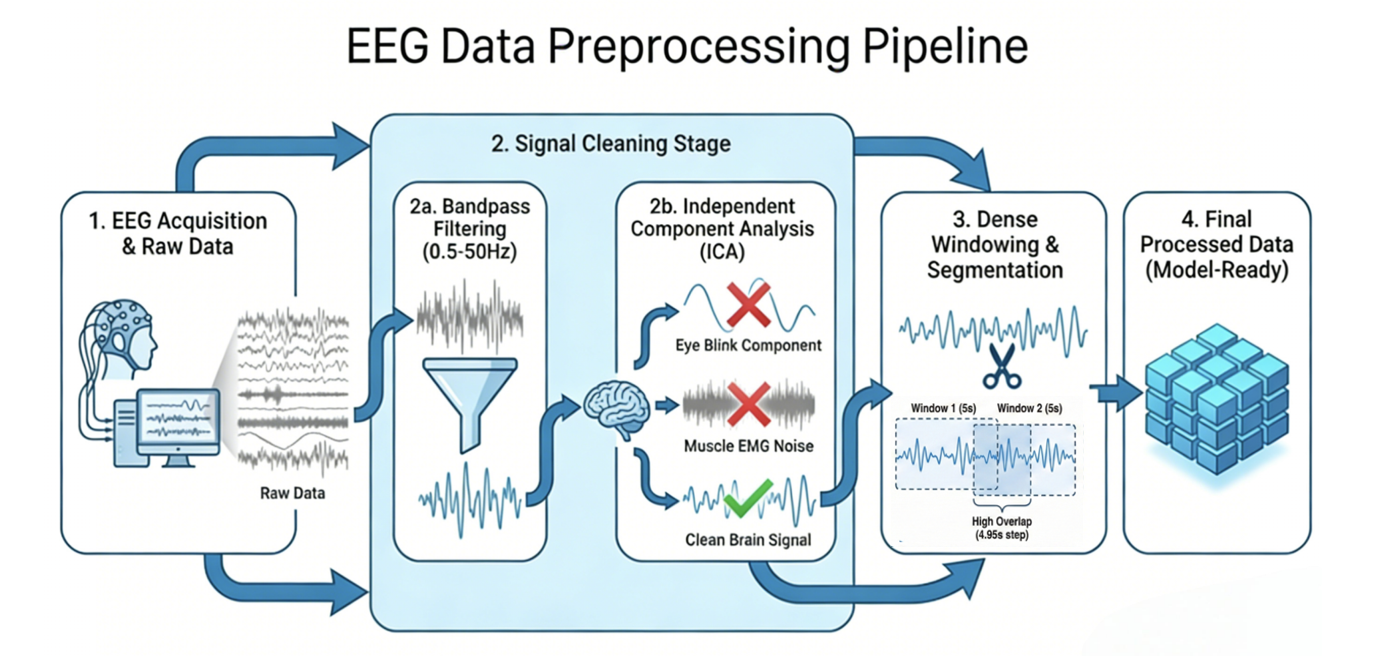}
\caption{Schematic overview of the unified EEG data preprocessing pipeline. The workflow follows a sequential progression across four stages: 
\textbf{(1) Data Acquisition:} Raw multichannel EEG signals containing physiological noise and baseline drift. 
\textbf{(2) Signal Cleaning:} A composite stage integrating band-pass filtering (0.5--50 Hz) and Independent Component Analysis (ICA). ICA decomposes signals into source components, allowing for the specific rejection of ocular  and myogenic artifacts while preserving neural activity. 
\textbf{(3) Dense Segmentation:} The cleaned continuous EEG is partitioned using a sliding-window strategy with high temporal overlap to maximize data utilization. 
\textbf{(4) Standardization:} The final output consists of aligned, fixed-length tensors ready for model ingestion. }
\label{fig:preprocessing_pipeline}
\end{figure*}

\section{Problem Formulation and Materials} \label{sec:problem}

\subsection{Problem Formulation}
    In this work, automated IED analysis is formulated as a joint detection and report generation problem under an evidence-grounded clinical decision support framework (Fig.~\ref{fig:task_overview}). Given a multichannel EEG recording segmented into short-time windows, each EEG segment is associated with a binary label indicating the presence or absence of IED. This label is determined by expert annotation or clinically verified event labels. EEG segments containing interictal epileptiform patterns are referred to as IED segments, while all remaining segments are regarded as non-IED segments. The primary objective is to determine whether a given EEG segment exhibits IED activity. Simultaneously, the framework aims to generate a clinically consistent textual EEG report that reflects the underlying electrophysiological characteristics. To enable interpretable decision-making, the problem is further constrained. Both detection and report generation must be grounded in retrieved historical EEG--text pairs that exhibit high semantic similarity to the query segment. Fig.~\ref{fig:task_overview} presents the proposed framework for joint IED detection and report generation. The input is a query EEG segment $X_q$. To ensure reliability, the system performs evidence-grounded analysis. It retrieves clinically relevant historical cases from an external knowledge base to support the inference. Based on this evidence, the framework achieves two objectives simultaneously: accurate IED detection (Objective 1) and interpretable report generation (Objective 2). The proposed formulation is evaluated on two complementary datasets: a private pediatric EEG dataset from Wuhan Children's Hospital and the public TUH EEG Events Corpus (TUEV).
    
    \subsection{Data Description}
    \subsubsection{Dataset 1}
    The dataset consisted of EEG recordings from 11 pediatric patients diagnosed with epilepsy, collected at Wuhan Children's Hospital. Detailed demographic and clinical information was available for each subject (see Table~\ref{tab:dataset1_profile}), including sex, age, epilepsy classification, disease duration, and seizure localization. EEG acquisition followed the international 10--20 electrode placement system, with a sampling frequency of 500 Hz to capture fine-grained neural dynamics during interictal and sleep states. Three experienced epileptologists independently annotated the onset and offset of IED, together with their waveform morphologies and spatial distributions. In addition to signal-level annotations, the experts produced structured clinical text descriptions for each EEG segment. These descriptions summarize the presence or absence of IED, waveform characteristics, anatomical localization, sleep-related modulation, and relevant clinical context. This process formed a paired EEG--text dataset. In total, 1206 IED events were identified, with substantial inter-subject variability in frequency, morphology, and duration, providing a heterogeneous and clinically representative dataset. The study protocol was approved by the Ethics Committee of Wuhan Children's Hospital (IRB No. 2022R034-E01).

    \subsubsection{Dataset 2}
    The dataset consisted of scalp EEG recordings from the TUEV (Version 2.0.3) \cite{ref30}, a publicly available benchmark derived from the Temple University Hospital EEG database. While Dataset 1 provides high-resolution data from a specific clinical center, TUEV is included to ensure a fair comparison and reproducibility on a standardized community benchmark. All recordings were acquired using standard clinical EEG systems with electrodes placed according to the international 10--20 system, ensuring consistency across sessions. The corpus contains precise onset and offset labels for multiple EEG event types. In this study, we specifically utilized segments annotating spike-and-slow-wave discharges (SPSWs), generalized periodic epileptiform discharges (GPEDs), and periodic lateralized epileptiform discharges (PLEDs) as IED samples, and background activity segments as non-IED samples. Segments corresponding to physiological artifacts or eye movements were excluded. The corpus is divided into disjoint training and evaluation sets, comprising 359 and 159 recordings, respectively. This collection offers substantial diversity in event morphology and recording conditions, providing a robust public benchmark for evaluating automated IED detection and interpretation methods.

\subsection{Preprocessing}
To ensure consistency across diverse data sources, a unified preprocessing pipeline was implemented for both Dataset~1 and Dataset~2 (Fig.~\ref{fig:preprocessing_pipeline}). \textbf{Signal processing:} Raw EEG was band-pass filtered (0.5--50~Hz) and resampled to 500~Hz. All recordings were mapped to a standardized 19-channel configuration following the international 10--20 system; missing channels in Dataset~2 were interpolated to match this montage. For Dataset~1, ICA was additionally applied to remove physiological artifacts (ocular and myogenic activity) while preserving neural components. \textbf{Segmentation and labeling:} Continuous recordings were segmented using a sliding window (5~s duration, 0.05~s stride) to densely sample transient IED patterns. Segments temporally overlapping with annotated epileptiform events were labeled as IED (positive), while the remaining segments were labeled as non-IED (negative). All steps were implemented using MATLAB 2018b and MNE-Python with consistent processing and labeling rules across datasets.

\textbf{Data partitioning:} For Dataset~1, stratified within-subject sampling was applied exclusively to construct the training set. This ensured robust representation of rare IED patterns during learning. All remaining segments were reserved for the held-out test set. Consequently, the test set contained approximately 100{,}000 segments ($N\approx 100{,}000$), preserving the natural clinical class imbalance ($\approx 32\%$ positive and $68\%$ negative). For Dataset~2, the official TUEV train--evaluation split was strictly followed. To evaluate morphological discrimination with reduced class-prior bias, we additionally formed balanced subsets by random subsampling within each split, resulting in an evaluation set of 40{,}000 segments (20{,}000 positive and 20{,}000 negative). Since TUEV does not provide clinician-authored reports, structured textual descriptions were generated using rule-based templates derived from the annotated event phenotypes and their spatial localization to enable multimodal EEG--text learning.

\section{Method} \label{sec:method}

    \begin{figure*}[t]
    \centering
    \includegraphics[width=0.98\textwidth]{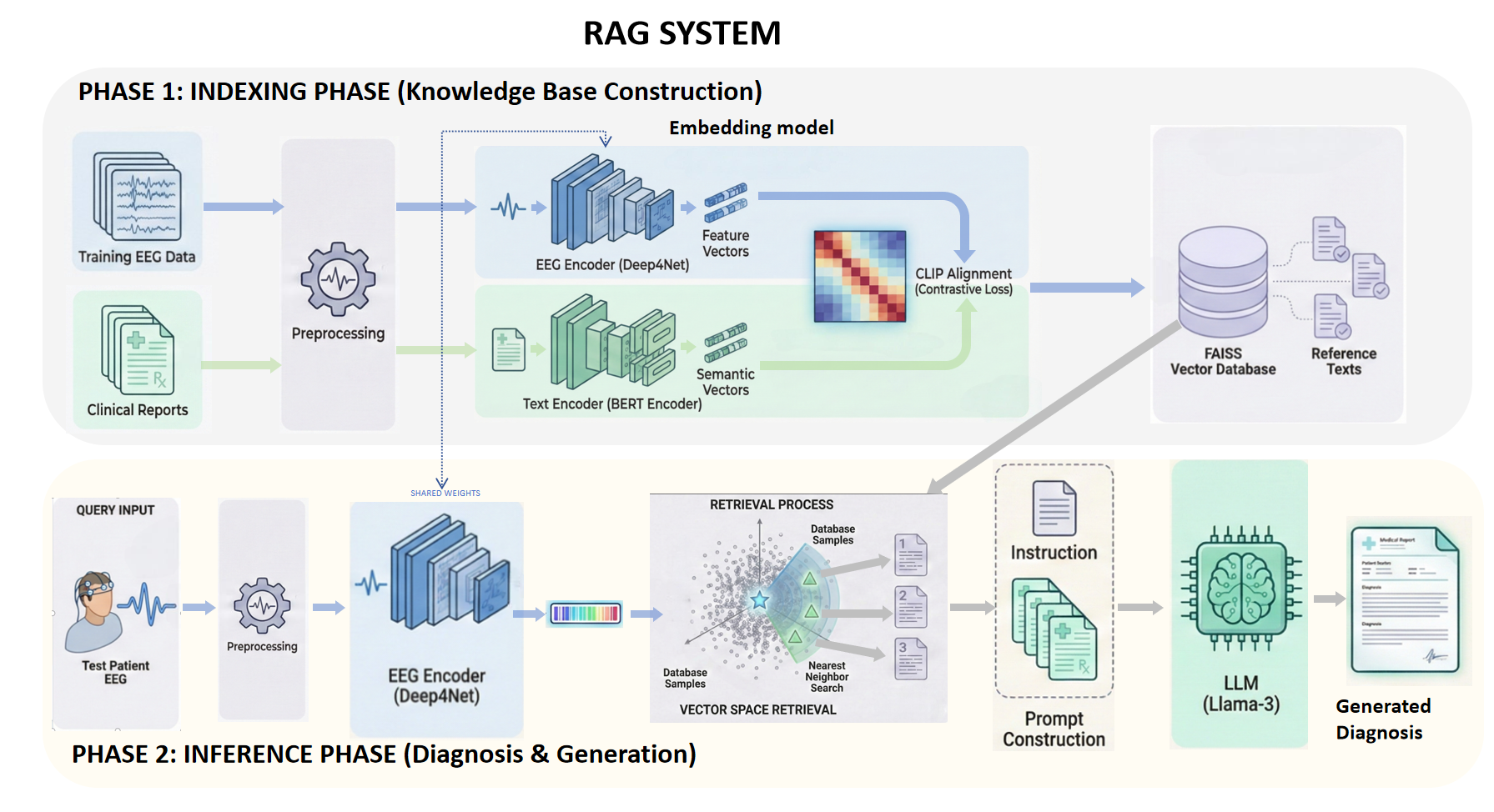}
    \caption{Overview of the proposed explainable multimodal RAG framework. The architecture follows a two-phase design: \textbf{(A) Indexing Phase} (top) and \textbf{(B) Inference Phase} (bottom). 
    \textbf{(A) Indexing:} Paired training EEG segments and expert-authored clinical reports are encoded by a dual-encoder model, where a Deep4Net-based EEG encoder and a BERT-based text encoder project multimodal inputs into a shared embedding space. The EEG embeddings are indexed in a FAISS vector database together with their associated report texts and labels, forming a searchable multimodal vector database. 
    \textbf{(B) Inference:} Given a query EEG segment, the same EEG encoder (shared weights) produces a query embedding, which retrieves the Top-$K$ most similar historical EEG cases via nearest-neighbor search (cosine similarity). The retrieved report texts are then assembled with task instructions into a retrieval-augmented prompt to condition an LLM, generating an evidence-grounded and interpretable EEG report.}
    \label{fig:framework}
    \end{figure*}

\subsection{Overall Framework}

As illustrated in Fig.~\ref{fig:framework}, the proposed framework operates in two distinct stages: Cross-Modal Representation Learning and Retrieval-Augmented Clinical Decision Support. In the first stage, a dual-encoder architecture is constructed. This architecture consists of an EEG encoder and a text encoder designed to map raw electrophysiological signals and corresponding clinical descriptions into a shared semantic latent space. Through the optimization of a contrastive objective, EEG representations are aligned with their semantic phenotypes. This alignment enables the construction of a searchable vector database where historical EEG cases are indexed as dense vectors.
    
    In the second stage, a query EEG segment retrieves its Top-$K$ nearest neighbors from the pre-computed index as clinically relevant evidence. The retrieved precedents are used for (i) vote-based evidence aggregation for detection and (ii) grounding an LLM to generate an evidence-based clinical EEG report. This design enables transparent, case-level interpretability and mirrors clinical case-based reasoning.

\subsection{Contrastive EEG--Text Semantic Alignment} \label{subsec:contrastive}
The goal of this module is to learn a shared embedding space for EEG segments and their paired clinical descriptions. This space supports similarity search and provides the basis for evidence retrieval.

\begin{figure*}[t]
    \centering
    \includegraphics[width=\textwidth]{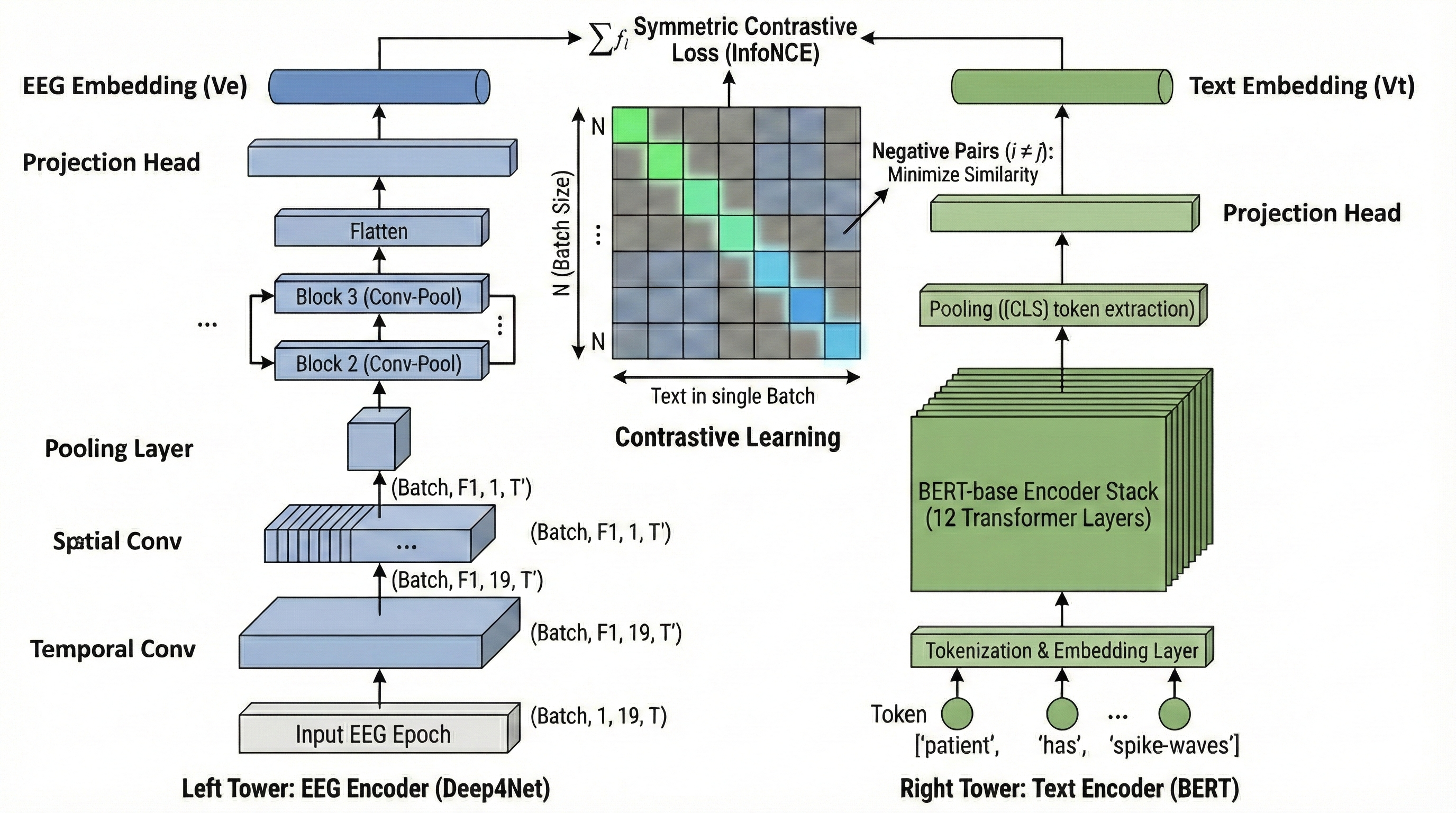}
    \caption{\textbf{Architecture of the Cross-Modal Contrastive Learning Framework.} The model employs a dual-tower structure to align electrophysiological signals with clinical narratives in a shared embedding space.
    \textbf{(Left) EEG Encoder (Deep4Net):} The encoder processes raw EEG inputs ($C=19, T=2500$) through a specialized hierarchy: 
    1) \textit{Temporal Convolution}: Extracts frequency features using $(1, 10)$ kernels, expanding feature depth to 25.
    2) \textit{Spatial Convolution}: Aggregates spatial information across all 19 channels using $(19, 1)$ kernels, compressing the spatial dimension to 1.
    3) \textit{Hierarchical Pooling}: Subsequent blocks progressively increase feature depth ($25 \to 200$) while reducing temporal resolution via Max Pooling. The final output is projected to a 512-dimensional embedding vector $\mathbf{v}_e$.
    \textbf{(Right) Text Encoder (BERT):} Clinical reports are tokenized and processed by a pre-trained \textit{BERT-base} model. The global semantic context is extracted via the \texttt{[CLS]} token and projected to a text embedding vector $\mathbf{v}_t \in \mathbb{R}^{512}$.
    \textbf{(Center) Joint Optimization:} The network is trained using a symmetric InfoNCE loss. The heatmap illustrates the objective: maximizing cosine similarity for matched positive pairs (diagonal, dark squares) while minimizing it for unmatched negative pairs (off-diagonal) within the batch.}
    \label{fig:contrastive_architecture}
\end{figure*}

We adopt Deep4Net as the EEG backbone encoder $f_\theta$~\cite{schirrmeister2017deep} and BERT-base-uncased as the text encoder $g_\phi$~\cite{devlin2019bert} (Fig.~\ref{fig:contrastive_architecture} and Table~\ref{tab:contrastive_config}). Both encoders are followed by a projection head and $L_2$ normalization. Under normalization, the inner product is equivalent to cosine similarity.

We train the dual encoder using the standard symmetric InfoNCE objective~\cite{oord2018representation,chen2020simple,radford2021learning}:
\begin{equation}
\begin{aligned}
\mathcal{L}_{\mathrm{con}}=\frac{1}{2N}\sum_{i=1}^{N}\Bigg[ &-\log\frac{\exp(\mathbf{z}_i^\top\mathbf{u}_i/\tau)}{\sum_{j=1}^{N}\exp(\mathbf{z}_i^\top\mathbf{u}_j/\tau)} \\
&-\log\frac{\exp(\mathbf{u}_i^\top\mathbf{z}_i/\tau)}{\sum_{j=1}^{N}\exp(\mathbf{u}_i^\top\mathbf{z}_j/\tau)} \Bigg],
\end{aligned}
\end{equation}
where $\mathbf{z}_i=\frac{f_\theta(X_i)}{\|f_\theta(X_i)\|_2}$ and $\mathbf{u}_i=\frac{g_\phi(Y_i)}{\|g_\phi(Y_i)\|_2}$.

\begin{table}[t]
\centering
\caption{Model architecture and training configuration.}
\label{tab:contrastive_config}
\begin{tabular}{ll}
\hline
\textbf{Component} & \textbf{Specification} \\
\hline
\multicolumn{2}{l}{\textbf{EEG Encoder Only}} \\
Input Shape & $X \in \mathbb{R}^{19 \times 2500}$ \\
Backbone & Deep4Net (4 conv blocks) \\
Channels & 25 $\rightarrow$ 50 $\rightarrow$ 100 $\rightarrow$ 200 \\
Pooling & MaxPool2D ($1 \times 3$) per block \\
\hline
\multicolumn{2}{l}{\textbf{Text Encoder Only}} \\
Backbone & BERT-base-uncased \\
Max Length & 77 tokens \\
Representation & \texttt{[CLS]} token \\
\hline
\multicolumn{2}{l}{\textbf{Shared Alignment}} \\
Projection Head & Linear $\rightarrow$ BN $\rightarrow$ ReLU $\rightarrow$ Dropout $\rightarrow$ Linear \\
Output Dim ($d$) & 512 \\
Loss Function & InfoNCE (Temperature $\tau$ learnable) \\
Optimizer & AdamW ($\beta_1=0.9, \beta_2=0.999$) \\
Learning Rate & $1 \times 10^{-4}$ \\
Batch Size & 128 \\
\hline
\end{tabular}
\end{table}

\subsection{Retrieval-Augmented Detection and Report Generation} \label{subsec:rag}
\begin{figure*}[t]
        \centering
        \includegraphics[width=\linewidth]{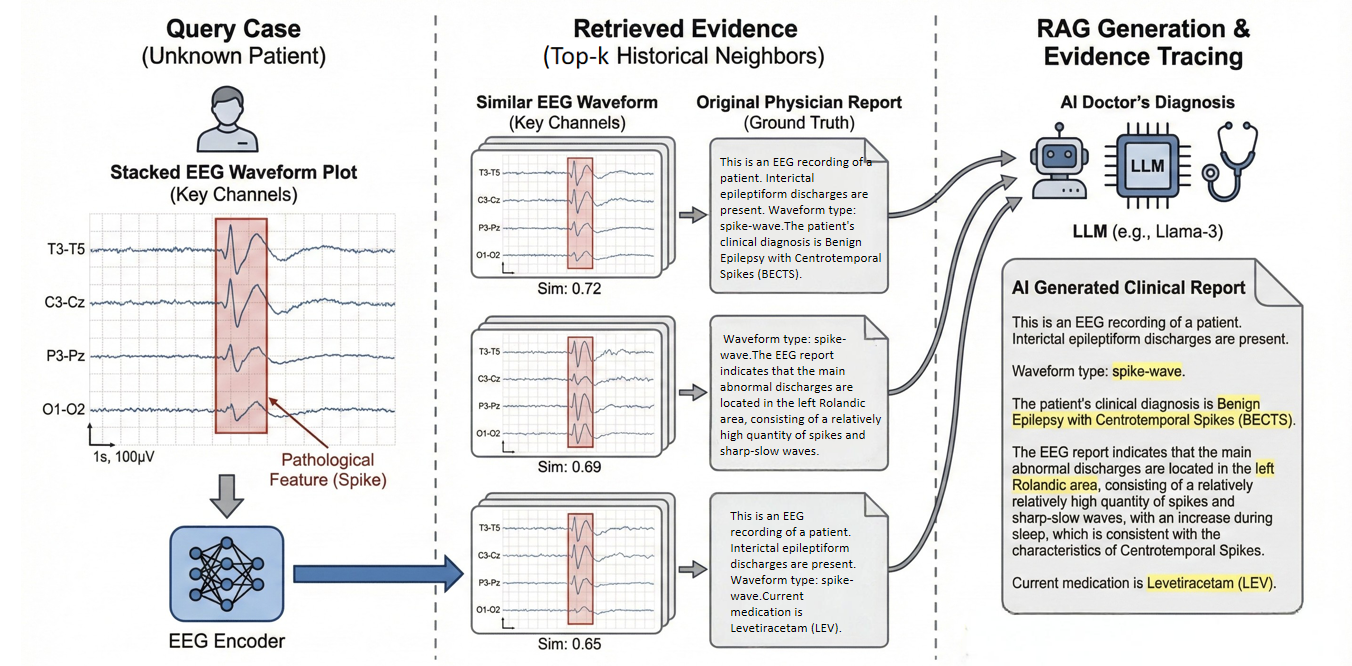}
        \caption{\textbf{Interpretability Case Study: Evidence-Grounded EEG Report Generation via Multimodal RAG.}
        \textbf{(Left) Input and query:} An EEG segment containing a suspected IED is provided as the query.
        \textbf{(Middle) Evidence retrieval:} Clinically relevant EEG--report pairs are retrieved from a FAISS-based vector database according to the learned EEG--text embedding similarity. The retrieved neighbors exhibit spike--wave morphology consistent with the query, and their associated clinician reports provide explicit diagnostic context.
        \textbf{(Right) Generation and traceability:} The final EEG report is generated by a large language model conditioned on the retrieved evidence, enabling case-based traceability by linking the generated statements to the retrieved clinical precedents.
        Overall, transparent evidence tracing is enabled by explicitly exposing retrieved neighbors and their original reports, thereby supporting interpretable IED detection and EEG report generation.}
        \label{fig:case_study}
    \end{figure*}

Retrieval-augmented inference supports both IED detection and evidence-grounded report generation (Fig.~\ref{fig:case_study}). We embed training EEG segments and index them using FAISS for fast inner-product search~\cite{johnson2019billion}. This setup follows the RAG paradigm~\cite{lewis2020retrieval}.

Let $\mathcal{D}_{\mathrm{train}}=\{(X_i,Y_i,\ell_i)\}_{i=1}^{M}$ denote the training corpus, where $\ell_i\in\{0,1\}$ is the ground-truth IED label. Given a query segment $X_q$, we compute its normalized embedding $\mathbf{z}_q$ and retrieve the Top-$K$ nearest neighbors by cosine similarity:
\begin{equation}
\mathcal{N}_K(X_q)=\arg\max_{i\in\{1,\dots,M\}}^{K}\ \mathbf{z}_q^\top \mathbf{z}_i,
\end{equation}
where $\mathbf{z}_q^\top\mathbf{z}_i$ is the inner product. Under $L_2$ normalization, inner-product search is equivalent to cosine similarity maximization.

\subsubsection{Evidence-Aggregated IED Detection}
We aggregate neighbor labels to obtain an evidence score. We define
\begin{equation}
E_q=\frac{1}{K}\sum_{j\in\mathcal{N}_K(X_q)}\ell_j,
\end{equation}
and predict $\hat{\ell}_q=\mathbb{I}(E_q\ge\gamma)$, where $\gamma\in(0,1]$ is a voting threshold.

Parameters $K$ and $\gamma$ were selected via grid search on a held-out validation split to maximize Balanced Accuracy.

\subsubsection{Reference-Guided EEG Report Generation via Deterministic, Constraint-Aware Prompting}
To improve factual consistency, we constrain report generation using retrieved references. We use Meta-Llama-3-8B-Instruct~\cite{dubey2024llama} as a deterministic selector and disable stochastic sampling. For each query, we inject a small set of Top-$K$ retrieved reports into the prompt context. The model outputs the exact text of the most relevant reference and must not edit or add any content. This constraint reduces unsupported statements and mitigates hallucinations~\cite{lewis2020retrieval}.

    \begin{table*}[t]
    \centering
    \caption{Constraint-aware prompt template for deterministic, reference-guided EEG report generation. The prompt enforces strict replication of a retrieved clinically verified reference to minimize unsupported content.}
    \label{tab:prompt_template}
    \renewcommand{\arraystretch}{1.15}
    \begin{tabular}{p{0.16\linewidth} p{0.80\linewidth}}
    \hline
    Field & Content / instruction template \\
    \hline
    System role &
    You are an expert EEG diagnostician. Your task is to output an EEG report that is identical to the most relevant reference provided below. \\
    \hline
    Context injection &
    Retrieved evidence from the vector database (Top-$K$ reports): \newline
    REFERENCE 1: [Report text of neighbor 1] \newline
    REFERENCE 2: [Report text of neighbor 2] \newline
    REFERENCE 3: [Report text of neighbor 3] \\
    \hline
    Constraints &
    (1) Select the most relevant reference from the provided context. \newline
    (2) Output the exact same text as the selected reference. \newline
    (3) Do not change any words, medical terms, punctuation, or formatting. \newline
    (4) Do not add any titles, labels, prefixes, notes, comments, or explanations. \newline
    (5) Copy the selected reference verbatim (word-for-word). \\
    \hline
    Output &
    One-paragraph EEG report (the verbatim reference text). \\
    \hline
    \end{tabular}
    \end{table*}

    For segments predicted as non-IED, a standardized normal statement was generated. A normal reference template was retrieved from the database, and the deterministic constraint was applied. This design reflects the stereotyped nature of normal EEG descriptions. It ensures that all outputs follow consistent hospital-style phrasing. Furthermore, strict grounding in verified precedents is maintained. The same constraint-aware prompt template was applied to both abnormal and normal segments, with retrieved references serving as the sole source of content.

Concretely, we assemble the prompt using the fixed template in Table~\ref{tab:prompt_template}. We then decode deterministically and output the selected reference verbatim.

\subsection{Evaluation Metrics}
\label{subsec:metrics}
We evaluate the framework from three aspects: (1) IED detection, (2) retrieval quality, and (3) report generation. We report confusion-matrix-based metrics for detection, ranking-based metrics for retrieval, and text-overlap metrics for report generation.

\subsubsection{IED Detection Metrics}
We report Balanced Accuracy (BA) and Weighted F1-score (WF1) as primary metrics. We also report Weighted Precision (WP) and Weighted Recall (WR). BA is defined as
\begin{equation}
\mathrm{BA}=\frac{1}{2}\left(\frac{TP}{TP+FN}+\frac{TN}{TN+FP}\right).
\end{equation}

\subsubsection{Retrieval Quality Metrics}
We evaluate retrieval relevance at $K\in\{1,2,3\}$, which matches the number of references injected into the prompt. A neighbor at rank $r$ is relevant if its label matches the query label. We report Precision@$K$ and Hit@$K$. Precision@$K$ is
\begin{equation}
P@K(q)=\frac{1}{K}\sum_{r=1}^{K} \mathbb{I}(\ell_{j_r}=\ell_q).
\end{equation}
We define Hit@$K$ as whether at least one relevant neighbor appears in the Top-$K$ list. We also report MAP and MRR to summarize ranking quality.

\subsubsection{Report Generation Metrics}
Report generation quality was evaluated by comparing the generated report $\hat{Y}_q$ with the expert-authored reference report $Y_q$ using standard natural language generation metrics. BLEU \cite{papineni2002bleu} was used to measure $n$-gram precision with a brevity penalty,
\begin{equation}
\mathrm{BLEU} = BP \cdot \exp\!\left( \sum_{n=1}^{4} w_n \log p_n \right),
\end{equation}
where $p_n$ denotes modified $n$-gram precision, $w_n$ are typically uniform weights satisfying $\sum_{n=1}^{4} w_n = 1$, and $BP$ is the brevity penalty. The reported $\mathrm{BLEU}$ was obtained by averaging BLEU scores across all evaluated queries.

ROUGE metrics \cite{lin2004rouge} were used to quantify lexical overlap between generated and reference reports. ROUGE-$n$ ($n \in \{1,2\}$) was computed as $n$-gram recall,
\begin{equation}
\mathrm{ROUGE}\text{-}n =
\frac{\sum_{g \in \mathcal{G}_n} \min\!\left(\mathrm{Count}_{\hat{Y}}(g), \mathrm{Count}_{Y}(g)\right)}
{\sum_{g \in \mathcal{G}_n} \mathrm{Count}_{Y}(g)},
\end{equation}
where $\mathcal{G}_n$ denotes the set of $n$-grams in the reference report $Y$. In addition, ROUGE-L was computed based on the longest common subsequence (LCS). Let $LCS(\hat{Y},Y)$ denote the length of the LCS between the generated and reference texts and $|Y|$ denote the length of the reference text. ROUGE-L recall was defined as
\begin{equation}
\mathrm{ROUGE}\text{-}L = \frac{LCS(\hat{Y},Y)}{|Y|}.
\end{equation}
The reported averages $\mathrm{ROUGE}\text{-}1$, $\mathrm{ROUGE}\text{-}2$, and $\mathrm{ROUGE}\text{-}L$ were obtained by averaging the corresponding ROUGE scores over all evaluated queries.

\section{Results} \label{sec:results}

\begin{table}[t]
\centering
\caption{Comparative performance of IED Detection on Dataset 1 and Dataset 2. Metrics are reported in \%.}
\label{tab:detection_performance}
\setlength{\tabcolsep}{5pt}
\renewcommand{\arraystretch}{1.2}
\resizebox{\columnwidth}{!}{
\begin{tabular}{llcccc}
\toprule
\multirow{2}{*}{\textbf{Dataset}} & \multirow{2}{*}{\textbf{Method}} & \multicolumn{4}{c}{\textbf{IED Detection Metrics}} \\
\cmidrule(lr){3-6}
& & \textbf{W-F1} & \textbf{B-Acc} & \textbf{W-Prec} & \textbf{W-Rec} \\
\midrule
\multirow{3}{*}{\textbf{Dataset 1}}
& NeuroLM & 64.66 & 67.35 & 72.13 & 63.54 \\
& IED-RAG (No-Text) & 83.45 & 83.78 & 84.90 & 83.04 \\
& \textbf{IED-RAG} & \textbf{88.81} & \textbf{89.17} & \textbf{89.57} & \textbf{88.60} \\
\midrule
\multirow{3}{*}{\textbf{Dataset 2}}
& NeuroLM & 70.37 & 70.79 & 72.03 & 70.79 \\
& IED-RAG (No-Text) & 68.76 & 68.76 & 68.77 & 68.76 \\
& \textbf{IED-RAG} & \textbf{71.05} & \textbf{71.38} & \textbf{72.41} & \textbf{71.38} \\
\bottomrule
\end{tabular}
}
\vspace{2pt}
\footnotesize \textit{W-F1: Weighted F1-score; B-Acc: Balanced Accuracy. ``IED-RAG (No-Text)'' denotes EEG-only supervised retrieval without EEG--text contrastive alignment. Best results are highlighted in bold.}
\end{table}

\begin{table*}[t]
\centering
\caption{Retrieval quality assessment on Dataset 1 and Dataset 2. Metrics are reported in \%.}
\label{tab:retrieval_performance}
\renewcommand{\arraystretch}{1.2}
\setlength{\tabcolsep}{4pt}
\resizebox{\textwidth}{!}{
\begin{tabular}{lcccccccccc}
\toprule
\multirow{2}{*}{\textbf{Dataset}} & \multirow{2}{*}{\textbf{Method}} & \multicolumn{8}{c}{\textbf{Retrieval Metrics}} \\
\cmidrule(lr){3-10}
& & \textbf{MAP} & \textbf{MRR} & \textbf{HR@1} & \textbf{P@1} & \textbf{HR@2} & \textbf{P@2} & \textbf{HR@3} & \textbf{P@3} \\
\midrule
\multirow{2}{*}{\textbf{Dataset 1}}
& IED-RAG (No-Text) & 89.63 & 90.62 & 86.52 & 86.52 & 92.90 & 82.35 & 95.63 & 79.40 \\
& \textbf{IED-RAG} & \textbf{92.77} & \textbf{93.33} & \textbf{90.63} & \textbf{90.63} & \textbf{94.81} & \textbf{88.20} & \textbf{96.65} & \textbf{86.57} \\
\midrule
\multirow{2}{*}{\textbf{Dataset 2}}
& IED-RAG (No-Text) & 68.43 & 73.83 & 62.03 & 62.03 & 76.06 & 62.82 & 83.10 & 63.12 \\
& \textbf{IED-RAG} & \textbf{71.65} & \textbf{76.64} & \textbf{65.39} & \textbf{65.39} & \textbf{79.54} & \textbf{65.84} & \textbf{85.72} & \textbf{66.09} \\
\bottomrule
\end{tabular}
}
\vspace{2pt}
\footnotesize \textit{NeuroLM is excluded as it does not perform explicit evidence retrieval.}
\end{table*}

\begin{figure*}[t]
\centering
\includegraphics[width=\textwidth]{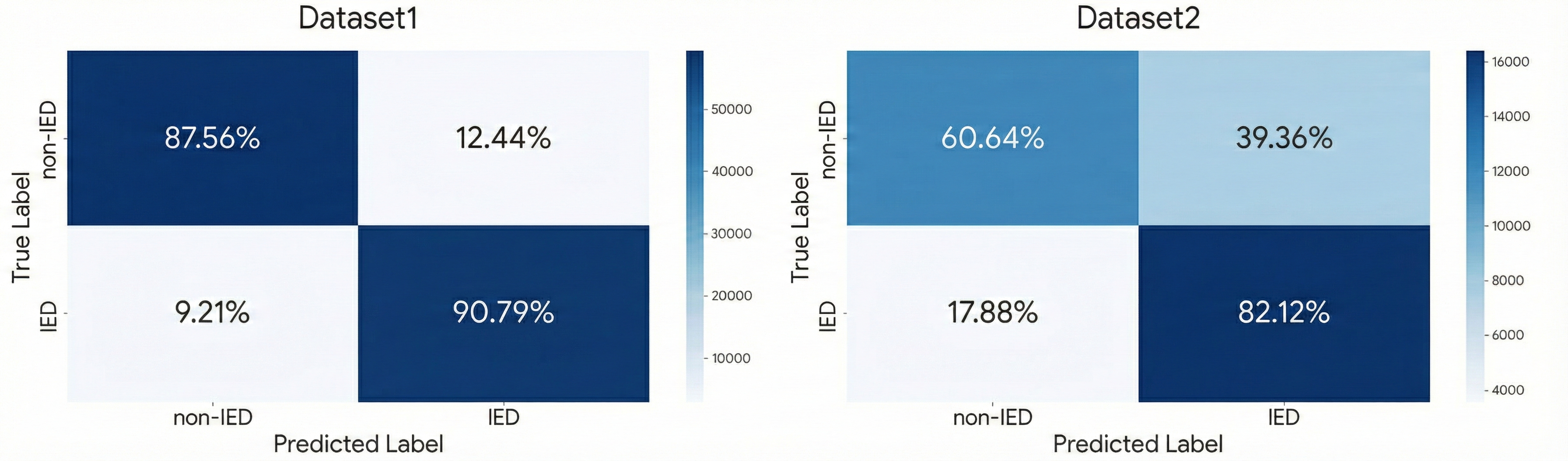}
\caption{Confusion matrices of the proposed model on the test sets of Dataset 1  and Dataset 2, reporting true negatives (TN), false positives (FP), false negatives (FN), and true positives (TP) for binary IED detection. }
\label{fig:confusion_matrix}
\end{figure*}

\subsection{IED Detection Performance}
The diagnostic performance of IED-RAG was evaluated on both Dataset 1 and Dataset 2. Binary classification (IED vs.\ non-IED) was performed via evidence aggregation over retrieved neighbors, and performance was summarized using Balanced Accuracy (BA), Weighted Precision (WP), Weighted Recall (WR), and Weighted F1-score (WF1), together with confusion-matrix counts ($TP$, $TN$, $FP$, $FN$) as shown in Fig.~\ref{fig:confusion_matrix}.

On Dataset 1, the framework achieved $\mathrm{BA}=89.17\%$, $\mathrm{WP}=89.57\%$, $\mathrm{WR}=88.60\%$, and $\mathrm{WF1}=88.81\%$. These values indicate strong screening performance in a controlled clinical environment.

On Dataset 2, the framework achieved $\mathrm{BA}=71.38\%$, $\mathrm{WP}=72.41\%$, $\mathrm{WR}=71.38\%$, and $\mathrm{WF1}=71.05\%$. These results reflect the increased difficulty of large-scale heterogeneous recordings compared with Dataset 1.

\subsection{Retrieval Quality Assessment}
A core premise of IED-RAG is that the retrieval module can identify historical cases semantically consistent with the query segment. This capability provides reliable evidence for downstream evidence aggregation and report generation. Retrieval quality was evaluated using ranking-based metrics defined in Section~\ref{subsec:metrics} (Table~\ref{tab:retrieval_performance}). These metrics included mean average precision (MAP), mean reciprocal rank (MRR), and the average hit rate and precision at cutoff depths $K\in\{1,2,3\}$ (HR@K and P@K). These depths match the number of references injected into the prompt context during inference.

On Dataset 1, strong retrieval relevance was observed. The system achieved $\mathrm{MAP}=92.77\%$ and $MRR=93.33\%$. At the top-ranked position, $HR@1=90.63\%$ and $P@1=90.63\%$, indicating that the nearest neighbor was frequently label-consistent with the query. As the retrieval depth increased, $HR@2=94.81\%$ and $HR@3=96.65\%$ were obtained, with corresponding precision values $P@2=88.20\%$ and $P@3=86.57\%$. Notably, an $HR@3$ of $96.65\%$ implies that the top-3 retrieved context window contained at least one label-consistent precedent for the majority of queries. This result confirms a reliable evidentiary basis for the subsequent retrieval-constrained report generation module.

On Dataset 2, retrieval was more challenging due to broader variability in artifacts and background patterns. The system achieved $\mathrm{MAP}=71.65\%$ and $MRR=76.64\%$. The top-ranked retrieval yielded $HR@1=65.39\%$ and $P@1=65.39\%$. Retrieval quality improved with deeper retrieval, reaching $HR@2=79.54\%$ and $HR@3=85.72\%$, with precision values $P@2=65.84\%$ and $P@3=66.09\%$. Overall, these results indicate that the vector database retrieves label-consistent precedents with high probability when multiple references are considered, even under increased signal heterogeneity. This robustness supports evidence-grounded inference and report standardization in downstream modules.

\subsection{Report Generation Performance}
This subsection presents the evaluation of generated EEG report quality. The assessment follows the deterministic, reference-guided generation protocol described in Section~\ref{subsec:rag}. Report quality was quantified by comparing the generated report $\hat{Y}_q$ with the expert-authored reference report $Y_q$. Text-overlap metrics defined in Section~\ref{subsec:metrics} were employed, including average BLEU and average ROUGE scores (ROUGE-1, ROUGE-2, and ROUGE-L). These results are summarized in Table~\ref{tab:report_generation}.

On Dataset 1, high textual consistency with the reference standards was observed. The framework achieved $\mathrm{BLEU}=89.61\%$, $\mathrm{ROUGE}\text{-}1=91.59\%$, $\mathrm{ROUGE}\text{-}2=91.09\%$, and $\mathrm{ROUGE}\text{-}L=91.58\%$. These results indicate that the generated reports closely match the expert-authored descriptions under the evidence-grounded replication constraint in a standardized clinical reporting environment.

On Dataset 2, report generation performance was lower. This outcome is consistent with the higher heterogeneity of multi-center recordings and greater variability in event phenotypes and report phrasing. The framework achieved $\mathrm{BLEU}=64.14\%$, $\mathrm{ROUGE}\text{-}1=75.57\%$, $\mathrm{ROUGE}\text{-}2=67.25\%$, and $\mathrm{ROUGE}\text{-}L=73.16\%$. The relatively lower $\mathrm{ROUGE}\text{-}2$ compared with $\mathrm{ROUGE}\text{-}1$ may reflect increased variation in local phrasing and word-order patterns across the corpus. Conversely, the $\mathrm{ROUGE}\text{-}1$ overlap indicates that core diagnostic terms and clinically relevant keywords are frequently preserved.

\section{Discussion} \label{sec:discussion}

\subsection{Comparison With NeuroLM Baseline}
    To benchmark the proposed Multimodal RAG framework against an established universal EEG foundation model, a comparative analysis was conducted using NeuroLM \cite{jiang2025neurolm}. NeuroLM follows a three-stage training paradigm. This paradigm consists of (1) vector-quantized (VQ) tokenization of EEG signals, (2) large-scale multi-channel autoregressive pre-training on EEG corpora, and (3) instruction tuning for downstream tasks. In this instruction-following inference setting, diagnostic outputs are primarily determined by implicit patterns encoded in model parameters. Consequently, explicit evidence grounding is absent.
    
    A pronounced performance gap was observed on Dataset 1. As summarized in Table~\ref{tab:detection_performance}, the proposed Multimodal RAG framework achieved a Balanced Accuracy of 89.17\% and a Weighted F1-score of 88.81\%, whereas NeuroLM achieved a Balanced Accuracy of 67.35\% and a Weighted F1-score of 64.66\%. Examination of the confusion matrices further indicated that NeuroLM produced substantially more false positives (FP=29{,}386) under the Dataset 1 setting. This result suggests that background activity and non-epileptiform patterns were frequently assigned to the IED-positive class. Such behavior could be attributed to domain mismatch and the absence of explicit, case-level grounding. Instruction-based inference was required to generalize to site-specific acquisition protocols and annotation criteria. In contrast, the proposed framework retrieved semantically similar EEG--report pairs from the local training distribution. This mechanism anchors the decision process to verifiable clinical precedents, thereby improving specificity and overall reliability.
    
    On Dataset 2, NeuroLM served as a strong baseline, achieving a Balanced Accuracy of 70.79\% and a Weighted F1-score of 70.37\% (Table~\ref{tab:detection_performance}). A modest yet consistent improvement was retained by the proposed framework, with a Balanced Accuracy of 71.38\% and a Weighted F1-score of 71.05\%. Although the numerical margin was smaller under the multi-institution setting, the results suggested that retrieval-augmented evidence did not introduce additional noise under in-the-wild conditions. Instead, a stable reference mechanism was provided. Furthermore, interpretability was simultaneously enabled through explicit evidence retrieval, a feature unavailable in purely parameter-based predictions.
    
    Overall, the comparison highlighted a fundamental distinction between implicit and explicit knowledge utilization. In NeuroLM, diagnostic patterns are required to be absorbed into model weights through large-scale pre-training and instruction tuning. This implicit knowledge may be vulnerable to domain shift when fine-grained, site-specific criteria are required. Conversely, the proposed Multimodal RAG framework decoupled knowledge storage (via the vector database) from reasoning and generation (via the large language model). Through contrastive EEG--text semantic alignment and explicit evidence retrieval, diagnostic decisions and generated EEG reports were grounded in retrievable clinical exemplars rather than unconstrained inference.

\begin{table*}[t]
\centering
\caption{Ablation Study on EEG Report Generation Quality. Metrics are reported in \%.}
\label{tab:report_generation}
\renewcommand{\arraystretch}{1.2} 
\setlength{\tabcolsep}{10pt} 
\begin{tabular}{llcccc}
\toprule
\multirow{2}{*}{\textbf{Dataset}} & \multirow{2}{*}{\textbf{Method}} & \multicolumn{4}{c}{\textbf{Generation Metrics}} \\
\cmidrule(lr){3-6}
& & \textbf{BLEU} & \textbf{ROUGE-1} & \textbf{ROUGE-2} & \textbf{ROUGE-L} \\
\midrule
\multirow{4}{*}{\textbf{Dataset 1}} 
& Random Retrieval (Baseline A) & 46.31 & 59.79 & 54.05 & 58.44 \\
& Random Retrieval (Baseline B) & 73.56 & 83.09 & 77.87 & 81.50 \\
& IED-RAG (No-Text) & 83.80 & 87.10 & 86.15 & 87.01 \\
& \textbf{IED-RAG} & \textbf{89.61} & \textbf{91.59} & \textbf{91.09} & \textbf{91.58} \\
\midrule
\multirow{4}{*}{\textbf{Dataset 2}} 
& Random Retrieval (Baseline A) & 50.19 & 65.70 & 54.65 & 62.04 \\
& Random Retrieval (Baseline B) & 61.79 & 74.03 & 65.04 & 71.56 \\
& IED-RAG (No-Text) & \textbf{64.77} & \textbf{75.99} & \textbf{68.00} & \textbf{73.45} \\
& \textbf{IED-RAG} & 64.14 & 75.02 & 66.39 & 73.16 \\
\bottomrule
\multicolumn{6}{l}{\footnotesize \textit{ Baseline A: Pure random retrieval; Baseline B: Label-matched random retrieval.}} \\
\end{tabular}
\end{table*}

\subsection{Semantic Retrieval vs.\ Random Retrieval}
    
    A central claim of the proposed framework is that interpretability arises from \emph{evidence grounding}. Each predicted decision and each generated report can be traced to retrieved historical EEG--text precedents that are semantically aligned with the query. To verify that this interpretability benefit is attributable to \emph{semantic retrieval} rather than to the language model itself, two negative-control baselines were constructed. In these baselines, semantic retrieval was replaced with randomized retrieval while the rest of the pipeline remained unchanged.
    
    \subsubsection{Experimental setup and controls}
    Two randomized retrieval baselines were considered. In Random-A, reference reports were sampled uniformly at random from the training corpus. This process destroys any semantic relationship between the query EEG and the retrieved evidence. In Random-B, reference reports were sampled randomly \emph{within the same predicted class label} (IED vs.\ non-IED). This strategy yields a stronger and more conservative baseline. It preserves global label priors while removing morphology- and localization-level alignment.
    
    Importantly, the classification decision rule was held constant across all conditions. Neighbor labels used for evidence aggregation were still obtained via the original FAISS-based semantic search. Only the textual evidence injected into the prompt was randomized. Under all conditions, the same constraint-aware prompting protocol was applied. Furthermore, the language model was used deterministically to replicate a selected reference report verbatim. Consequently, differences in text-overlap metrics directly reflect the semantic appropriateness of the retrieved evidence, rather than generative variability.
    
    \subsubsection{Quantitative impact on evidence-grounded reporting}
    Replacing semantic retrieval with random retrieval caused a substantial degradation in report overlap metrics (see Table~\ref{tab:report_generation}). On Dataset 1, semantic retrieval achieved $\mathrm{BLEU}=89.61\%$ and $\mathrm{ROUGE}\text{-}L=91.58\%$. In contrast, Random-A dropped sharply to $\mathrm{BLEU}=46.31\%$ and $\mathrm{ROUGE}\text{-}L=58.44\%$. Random-B partially recovered performance ($\mathrm{BLEU}=73.56\%$, $\mathrm{ROUGE}\text{-}L=81.50\%$) but remained substantially below semantic retrieval.
    
    A similar trend was observed on Dataset 2. Semantic retrieval achieved $\mathrm{BLEU}=64.14\%$ and $\mathrm{ROUGE}\text{-}L=73.16\%$, while Random-A decreased to $\mathrm{BLEU}=50.19\%$ and $\mathrm{ROUGE}\text{-}L=62.04\%$. Random-B again produced intermediate scores ($\mathrm{BLEU}=61.79\%$, $\mathrm{ROUGE}\text{-}L=71.56\%$).
    
    \subsubsection{Interpretability implications}
    These results highlight an important distinction between \emph{lexical similarity} and \emph{evidence relevance}. Clinical EEG reports are often highly templated and class-dependent. Therefore, retrieving any report from the correct label category can yield deceptively high BLEU or ROUGE scores, even when the retrieved case is unrelated to the query. In this sense, Random-B produces reports that are stylistically plausible yet insufficiently grounded at the case level.
    
    In contrast, semantic retrieval enforces fine-grained alignment in the learned embedding space. This process retrieves precedents that share specific morphological and spatial characteristics with the query signal. Report generation is constrained to replicate retrieved precedents deterministically. Therefore, semantic retrieval ensures that the generated report is not only traceable to an existing record but also justified by a clinically relevant analog. These findings demonstrate that semantic retrieval is essential for explainability beyond label correctness. It enables evidence-based reasoning rather than template-level imitation.

\begin{figure*}[t]
\centering
\includegraphics[width=\textwidth]{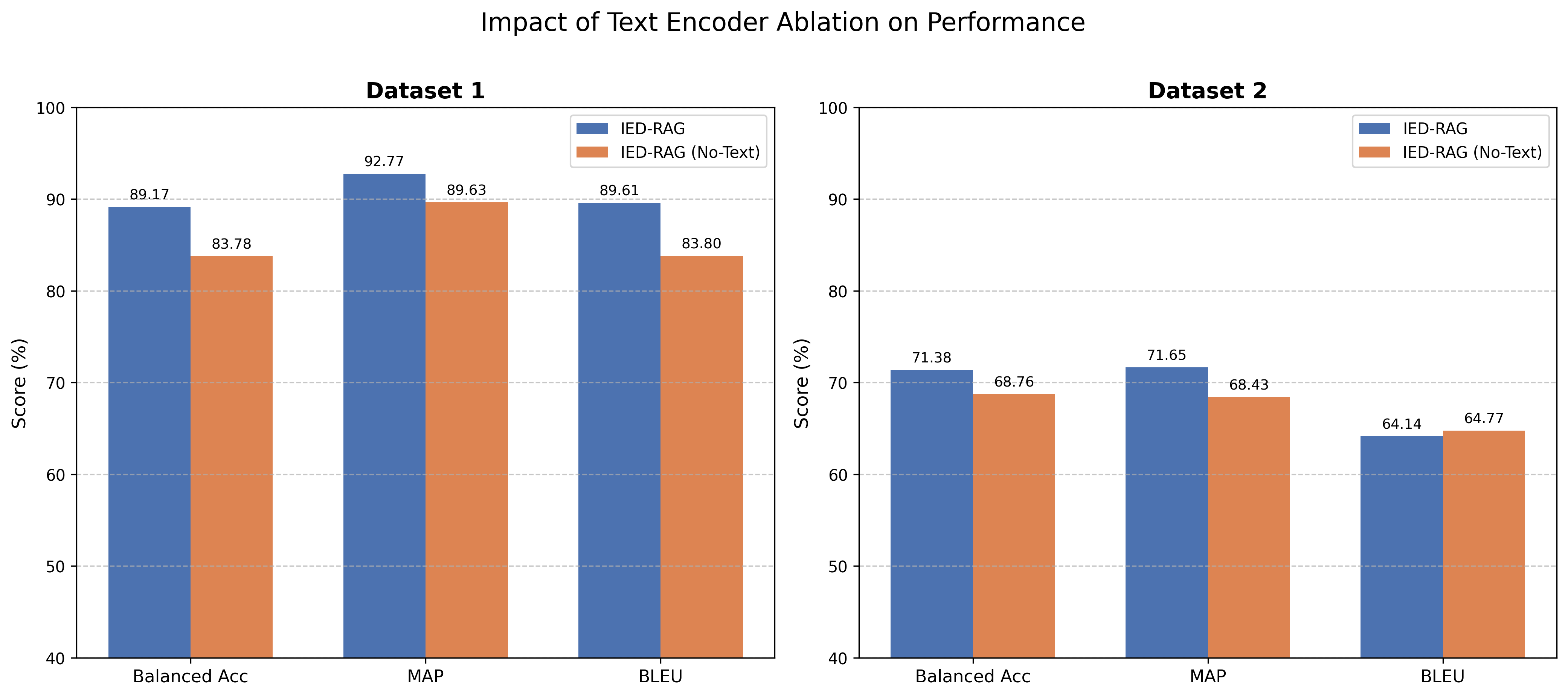}
\caption{Impact of Text Encoder Ablation on Performance. The grouped bar chart compares the proposed IED-RAG framework (Blue) versus the No-Text ablation baseline (Orange) across three evaluation dimensions: IED Detection (Balanced Accuracy), Evidence Retrieval (MAP), and Report Generation (BLEU). A consistent performance degradation is observed upon removing the text encoder, validating the necessity of semantic alignment. Note that on Dataset 2, BLEU shows a slight increase due to the generic nature of retrieval without semantic constraints, as discussed in the text.}
\label{fig:ablation_study}
\end{figure*}

\subsection{Ablation Study: Removing the Text Encoder}
    \label{subsec:ablation_no_text}
    To further elucidate the role of cross-modal semantic alignment in supporting interpretability, an ablation study was conducted. In this study, the text encoder and the contrastive EEG--text alignment objective were removed. In this variant, the EEG encoder was trained purely under a supervised classification objective. The resulting EEG embeddings were subsequently reused to construct the vector database for retrieval and downstream report generation. This setting preserves the overall retrieval--aggregation--generation pipeline. However, explicit semantic grounding between EEG signals and clinical language is eliminated.
    
    \subsubsection{Ablation design}
    Concretely, the contrastive learning stage described in Section~\ref{subsec:contrastive} was replaced with a standard supervised training regime. In this regime, the EEG encoder was optimized solely to predict the binary IED label using a linear classification head. After training, the classifier head was discarded. The penultimate EEG embeddings were extracted, $L_2$-normalized, and indexed using FAISS in the same manner as the full model. Retrieval, evidence aggregation, and constraint-aware report generation were then performed identically to the proposed framework. This configuration ensures a fair comparison. As a result, any observed differences can be attributed specifically to the absence of EEG--text semantic alignment rather than to changes in inference logic or prompting.

\subsubsection{Quantitative impact on detection, retrieval, and reporting}
    Removing the text encoder led to a consistent degradation across all three evaluation dimensions (Fig.~\ref{fig:ablation_study}). On Dataset 1, Balanced Accuracy dropped from $89.17\%$ to $83.78\%$, and the Weighted F1-score decreased from $88.81\%$ to $83.45\%$. Similar declines were observed on Dataset 2, where Balanced Accuracy decreased from $71.38\%$ to $68.76\%$ and Weighted F1-score from $71.05\%$ to $68.76\%$. These results indicate that embeddings learned solely for discrimination are less effective for neighbor-based evidence aggregation.
    
    Retrieval quality also deteriorated (Table~\ref{tab:retrieval_performance}). On Dataset 1, MAP decreased from $92.77\%$ to $89.63\%$, and $MRR$ dropped from $93.33\%$ to $90.62\%$. On Dataset 2, MAP declined from $71.65\%$ to $68.43\%$, with a corresponding reduction in hit rates at all evaluated depths. This result suggests that, without textual supervision, the learned embedding space prioritizes class separability over fine-grained morphological similarity. Consequently, the relevance of retrieved precedents is reduced. This degradation in retrieval relevance further propagates to the report generation stage.
    
    Finally, report generation quality declined (Table~\ref{tab:report_generation}), highlighting a limitation of purely supervised metrics. On Dataset 1, $\mathrm{BLEU}$ decreased from $89.61\%$ to $83.80\%$. On Dataset 2, BLEU exhibited a modest increase ($64.14\% \rightarrow 64.77\%$). This outcome can be attributed to the ablated model frequently retrieving generic normal templates. Such templates share high lexical overlap with the reference text despite lower diagnostic precision. A disconnect is observed between the reduction in classification accuracy ($-2.62\%$) and the stability of lexical metrics. This finding underscores that lexical similarity alone does not guarantee faithful physiological grounding.
    
    \subsubsection{Implications for interpretability}
    This ablation highlights a key distinction between \emph{discriminative} and \emph{interpretable} representations. Supervised training enables the EEG encoder to separate classes effectively. However, it does not enforce alignment between specific electrophysiological patterns and their linguistic descriptions. Consequently, retrieval becomes label-driven rather than morphology-driven. This shift weakens the explanatory link between the query EEG and the retrieved evidence.
    
    In contrast, the full cross-modal alignment model explicitly couples EEG features with clinical language. This coupling ensures that retrieved neighbors are not only label-consistent but also semantically meaningful at the level of waveform morphology and spatial localization. Therefore, the observed performance gap supports the central claim of this work. Explicit EEG--text semantic alignment is essential for transforming retrieval from a post-hoc lookup mechanism into a transparent, evidence-grounded explanation. Such explanations can be meaningfully inspected and trusted in clinical settings.

\section{Conclusion and future work} \label{sec:conclusion}
In this study, identifying clinically interpretable patterns from raw EEG signals remains a persistent challenge. To address this, we proposed IED-RAG, a multimodal Retrieval-Augmented Generation framework for explainable IED detection and report generation. By aligning EEG with clinical text in a shared semantic space and conditioning generation on retrieved evidence, the framework bridges the gap between black-box deep learning and case-based clinical reasoning. Experimental results on both private and public datasets demonstrate that explicit evidence retrieval not only enhances diagnostic performance but also ensures that generated reports are grounded in verifiable precedents. Future work will extend this paradigm to continuous long-term monitoring and broader event taxonomies, further integrating AI transparency into routine neurophysiological workflows.

\section*{DATA AVAILABILITY STATEMENT}
Data and code will be available upon request.

\end{document}